\documentclass[preprint,aps,prl,showpacs]{revtex4}
\usepackage{epsfig}
\begin{document}
\title{Island formation without attractive interactions.}
\author{A. P. J. Jansen}
\affiliation{Laboratory of Inorganic Chemistry and Catalysis, ST/SKA\\
  Eindhoven University of Technology\\
  P. O. Box 513, 5600 MB Eindhoven, The Netherlands}
\date{\today}
\begin{abstract}
  We show that adsorbates on surfaces can form islands even if there no
  attractive interactions. Instead strong repulsion between adsorbates
  at short distances can lead to islands, because such islands increase
  the entropy of the adsorbates that are not part of the islands. We
  suggest that this mechanism cause the observed island formation in
  O/Pt(111), but it may be important for many other systems as well.
\end{abstract}

\maketitle
%
%
Lateral interactions between adsorbates are extremely important for the
kinetics of surface reactions, mainly because they determine the
structure of the adlayer. Island formation is invariably assigned to
attractive interaction between the adsorbates, but reliable estimates
for lateral interactions are hard to obtain so that there is really
little know about such interactions. State-of-the-art calculations of
the lateral interactions for atoms and small molecules yield reliable
repulsive interactions, but attractive interactions, which are generally
weaker, are harder to obtain. Moreover, other experimental results may
only be consistent with repulsive interactions or attractive
interactions that are too small to stabilize islands. We discuss this
for oxygen atoms on Pt(111), which is a particularly well studied
system. We show that the islands that are observed for that system may
be formed without attractive interactions.  On the contrary, strong
repulsive interactions between adsorbates at short distance may lead to
islands because this lowers the entropy.

Low Energy Electron Diffraction (LEED) patterns of O/Pt(111) indicate
island formation with a $p(2\times 2)$ structure,\cite{par89b} which is
assigned to attractive interaction between oxygen atoms at a distance of
$2a$ with $a$ the distance between two neighboring fcc hollow adsorption
sites.\cite{tan04,nag05a,sen07a} Recently there have been
Density-Functional Theory (DFT) calculations of the lateral interactions
that showed indeed attraction at that distance.\cite{tan04,jan05} The
problem with DFT calculations is whether this interaction can be
calculated reliably. We have also done DFT calculations of a large
number of adlayer structures of O/Pt(111), but instead of determining
the lateral interaction by straightforward multivariate linear
regression we used a cross validation method as
well.\cite{jan02b,jan05,her06b} This is a statistical technique that has
been used extensively to determine the interactions between atoms in
alloys reliably,\cite{wal02,blu04} and also more recently for lateral
interactions.\cite{zha07c} We found that it was not possible to compute
an accurate interaction at distance $2a$. If we nevertheless tried to
do that, we found error bars that were almost an order of magnitude
larger than the absolute value of the interaction itself. If we varied
the set of adlayer structures, from which the lateral interactions were
determined, we found a large variation in the value of the interaction;
most of the time it was repulsive, but sometimes attractive. So we
concluded that a value of this interaction from DFT calculations is not
to be trusted.

That current DFT may not be able to say anything about the interaction
at the $2a$ distance does not mean that it can not be
attractive. However, the presence of absence of an attractive
interaction in O/Pt(111) has also been discussed by Zhdanov and Kasemo
while discussing Temperature-Programmed Desorption (TPD) experiments of
this system.\cite{zhd98c} In these spectra there is no indication that
there is an attractive lateral interaction. Their conclusions were that
if there is an attractive interaction then it so small that it can not
lead to island formation at the temperatures of the LEED experiment. We
have refined the kinetic Monte Carlo simulation of the TPD spectra that
were used by Zhdanov and Kasemo,\cite{jan02b} and determined the lateral
interactions by fitting the simulated TPD spectra to the experimental
ones. No attractive interactions were obtained. Moreover, the values we
obtained in this way agreed very well with those obtained from DFT
calculations when cross validation was used.

These observations lead to the question if it is possible to have island
formation without attractive interactions. We will show in this paper
that this is indeed possible, because of entropic reasons. Remarkably,
we will show that small islands lead to a higher entropy only when there
are repulsive interaction, not at the distances between the adsorbates
as observed in the island, but at shorter distances.

To study the effect of entropy on the island formation in O/Pt(111) we
have modeled the system as a hexagonal grid representing the fcc hollow
sites, which are the preferred sites of oxygen atoms.\cite{off06} The
interactions between the oxygen atoms are modeled using hard-sphere
interactions. These are such that two nearest-neighbor sites can not
both be occupied at the same time, and neither can two next-nearest
neighbor sites. The shortest distance between two oxygen atoms is then
$2a$. Apart from these two hard-sphere interactions there no other
interactions in our model. This model has been studied by Koper and
Lukkien to model the butterfly in voltammetry.\cite{kop00} It has a
order-disorder phase transition at around $0.18\,$ML. At higher
coverages the adlayer has the $p(2\times 2)$ structure also observed in
LEED. We are however interested at lower coverages where the adlayer has
no long-range order.

\begin{figure}[ht]
\begin{center}
\epsfig{figure=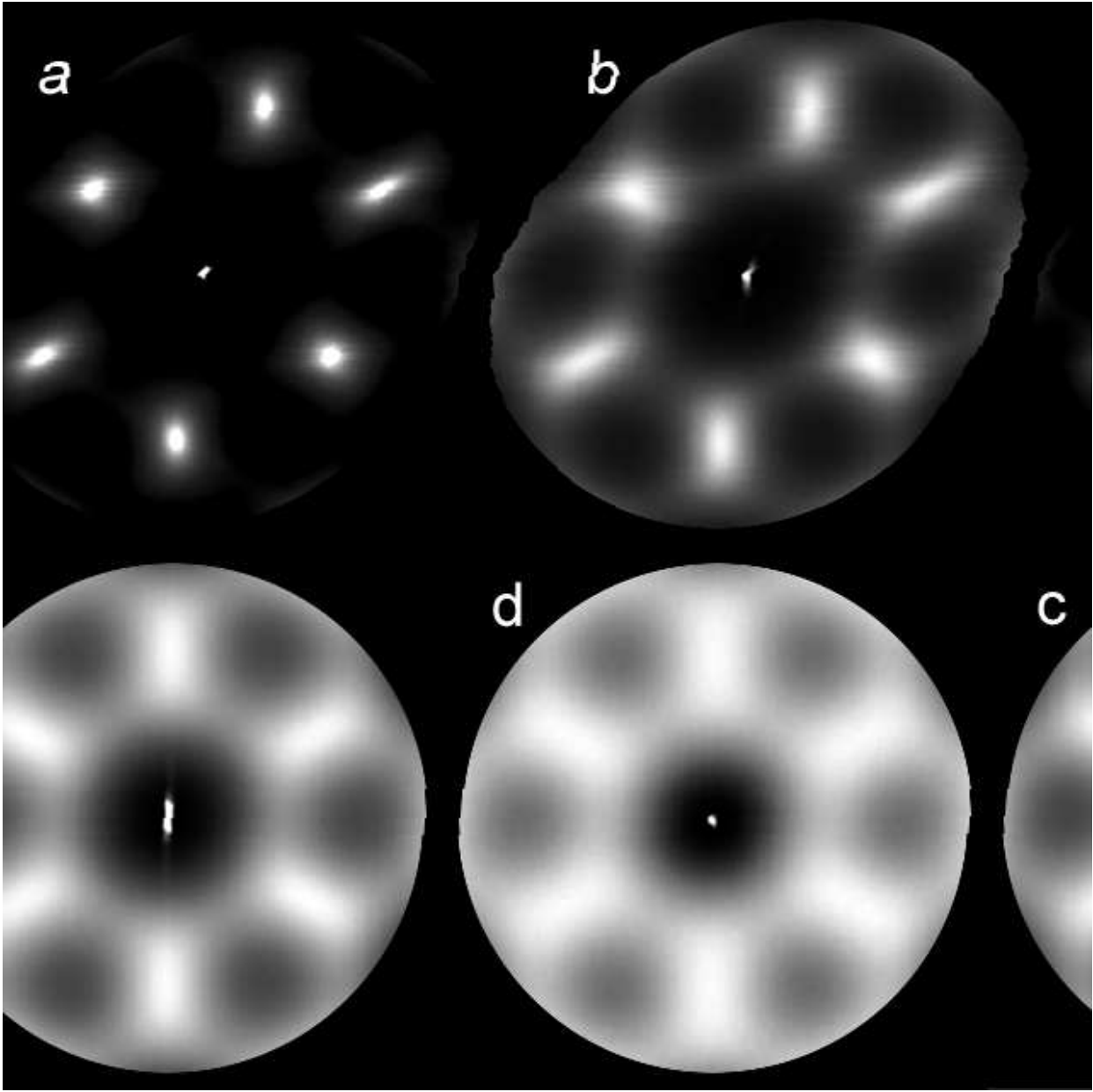,width=\hsize}
\caption{Simulated LEED patterns for the hard-sphere model of O/Pt(111)
  at coverages 0.16 (a), 0.12 (b), 0.08 (c), and 0.04 (d). The relative
  intensity of the peaks are 100, 11, 3.7, and 1.3, respectively.
  \label{fig:hsleed}}
\end{center}
\end{figure}

Fig.~\ref{fig:hsleed} shows simulated LEED patterns for a number of
coverages. It can clearly be seen that the adlayer has some structure.
At coverages just below the order-disorder transition the peaks are
sharp. The pattern is characteristic for a $p(2\times 2)$ structure with
a nearest distance between the adsorbates of $2a$. At very low coverages
the peaks become diffuse, but are still clearly visible.

To understand how the LEED pattern arises it is convenient to look at
another one-dimensional model that has essential the same
characteristics as the hard-sphere model for O/Pt(111). In this model we
have $S$ sites numbered from $0$ to $S-1$ with periodic boundary
conditions. There are $A$ adsorbates, and one adsorbate is always
adsorbed on site $0$. Two neighboring sites can not both be occupied by
an adsorbate at the same time. The probability that a site $n$ is
occupied, $P(n)$, can be determined from the fundamental hypothesis of
statistical mechanics that all acceptable configurations are equally
likely. Because two adsorbates can not be nearest neighbors, we have
$P(1)=P(S-1)=0$. The probability $P(2)$ is equal to the ratio of the
number of configurations with site 2 occupied and the total number of
configuration. This is given by
\begin{equation}
  P(2)
  ={N(A-2,S-5)\over N(A-1,S-3)}
\end{equation}
where $N(n,L)$ is the number of configurations with $n$ adsorbates
distributed over $L$ consecutive sites. We are assuming that the
adsorbates are indistinguishable, so that we have the recursion relation
\begin{equation}
  N(n,L)=N(n,L-1)+N(n-1,L-2)
\end{equation}
with the boundary conditions $N(n,L)=\delta_{n0}$ if $L\le 0$. (The
expressions change somewhat when we assume distinguishable adsorbates,
but the probabilities $P(n)$ remain the same.)

For $A=2$ we have $N(0,L)=1$ and $N(1,L)=L$ from the recursion relation
so that $P(2)=1/(S-3)$. This value is equal to the average occupation of
sites 2 to $S-2$. This is to be expected; the second adsorbate will have
equal probability to occupy any of the sites 2 to $S-2$. For $A=3$ we
have $N(2,L)=(L-1)(L-2)/2$ if $L\ge 3$ so that $P(2)=2/(S-4)$. We see
that in this case the probability of occupation is more than double the
average occupation, which is the first indication that there is a
tendency for clustering.

If site 3 is occupied, then the other adsorbates must be somewhere at
sites 5 to $S-2$. This means
\begin{equation}
  P(3)
  ={N(A-2,S-6)\over N(A-1,S-3)}.
\end{equation}
For $A=2$ we get $P(3)=P(2)=1/(S-3)$. For $A=3$ we get
$P(3)=2(S-6)/((S-4)(S-5))$. Again we have a higher probability than the
average occupation if $S\ge 7$, but $P(3)=0$ if $S=6$. In any case we
have $P(2)>P(3)$. We see that there is a tendency for the other
adsorbates to be as close as possible to the adsorbate that is always at
site 0. The tendency lessens if one goes farther from site 0. When $A=3$
and $S=6$ we have $P(3)=0$, because there is a maximum number of
adsorbates with alternating sites occupied and vacant.

\begin{figure}[ht]
\begin{center}
\epsfig{figure=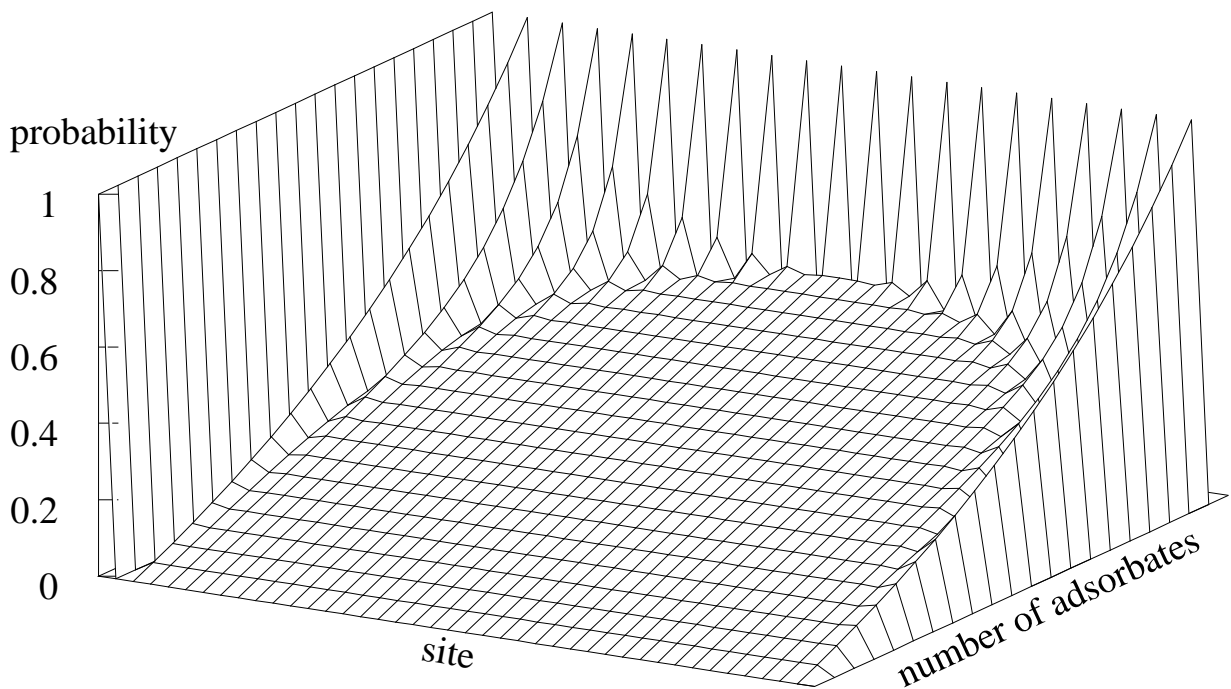,width=\hsize}
\caption{Occupation of all sites for the one-dimensional model with
  42 sites. The number of adsorbates ranges from 1 to 21 being small
  in front and large at the back. Site 0 on the left is always occupied.
  \label{fig:onedim}}
\end{center}
\end{figure}

If the number of adsorbates increases and when we look at the occupation
of sites for farther from site 0, then the analytical expressions for
the probabilities of occupation become quite complex. It is possible to
show that
\begin{eqnarray}
  P(n)&=&{1\over N(A-1,S-3)}\\
  &\times&\sum_{m=0}^{A-1} N(m,n-3)N(n-m-1,S-m-4).\nonumber
\end{eqnarray}
It is hard to see from this expression how $P(n)$ varies with $n$. It
seems therefore easier for moderate values of $S$ to simply do a
simulation in which all configurations are generated, and from that
determine the occupation of all sites. Fig.~\ref{fig:onedim} shows the
result for $S=42$. The tendency, which we mentioned already, of the
adsorbates to cluster is apparent from this figure even if there are
only a few adsorbates. We also see that we get ``islands'' with
adsorbates separated by a distance that is twice the distance between
neighboring sites. The origin of this clustering is the fact that when
two subsequent adsorbates are closer together, then the other adsorbates
have more sites over which to distribute, and hence a higher entropy.

\begin{figure}[ht]
\begin{center}
\epsfig{figure=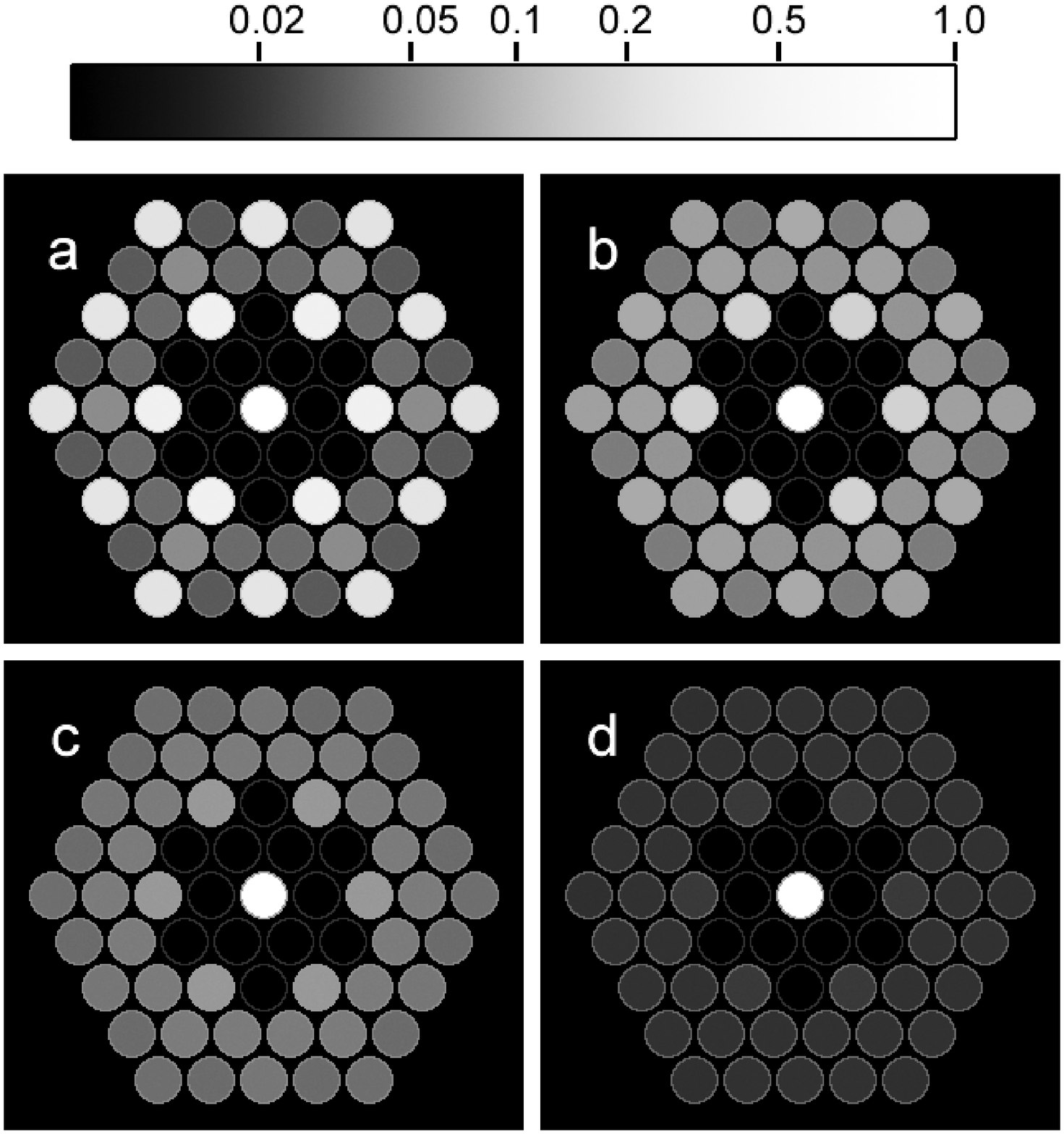,width=\hsize}
\caption{Probability of occupation of sites for the hard-sphere model of
  O/Pt(111). The central site is occupied in each case. The brightness
  indicates the probability that neighboring sites are occupied (see the
  scale at the top). The coverages are 0.16 (a), 0.12 (b), 0.08 (c), and
  0.04 (d).
  \label{fig:hsopt111}}
\end{center}
\end{figure}

The same probabilities can be computed for the hard-sphere model of
O/Pt(111) (see Fig.~\ref{fig:hsopt111}). This can either be done with a
small grid and generating all possible configurations, or with a larger
grid and doing a Monte Carlo simulation. We did both using a $10\times
10$ grid for generating all configurations, and a $64\times 64$ grid for
Monte Carlo simulations. The results were the same.

For coverages just below the phase transition there is a clear
indication already that a $p(2\times 2)$ structure is being formed (see
Fig.~\ref{fig:hsopt111}a). For lower coverages this structure becomes
harder to see in the figure. (The nearest and next-nearest sites are not
occupied, of course.) There is however a higher probability than average
for the next-next-nearest neighbor site; i.e., the nearest site to be
occupied in a $p(2\times 2)$ structure. This is clearly visible at a
coverage of 0.12, also visible at 0.08, but hard to see at 0.04,
although the occupation of the next-next-nearest neighbor site is still
about 11\% higher than average even at this low coverage.

\begin{figure}[ht]
\begin{center}
\epsfig{figure=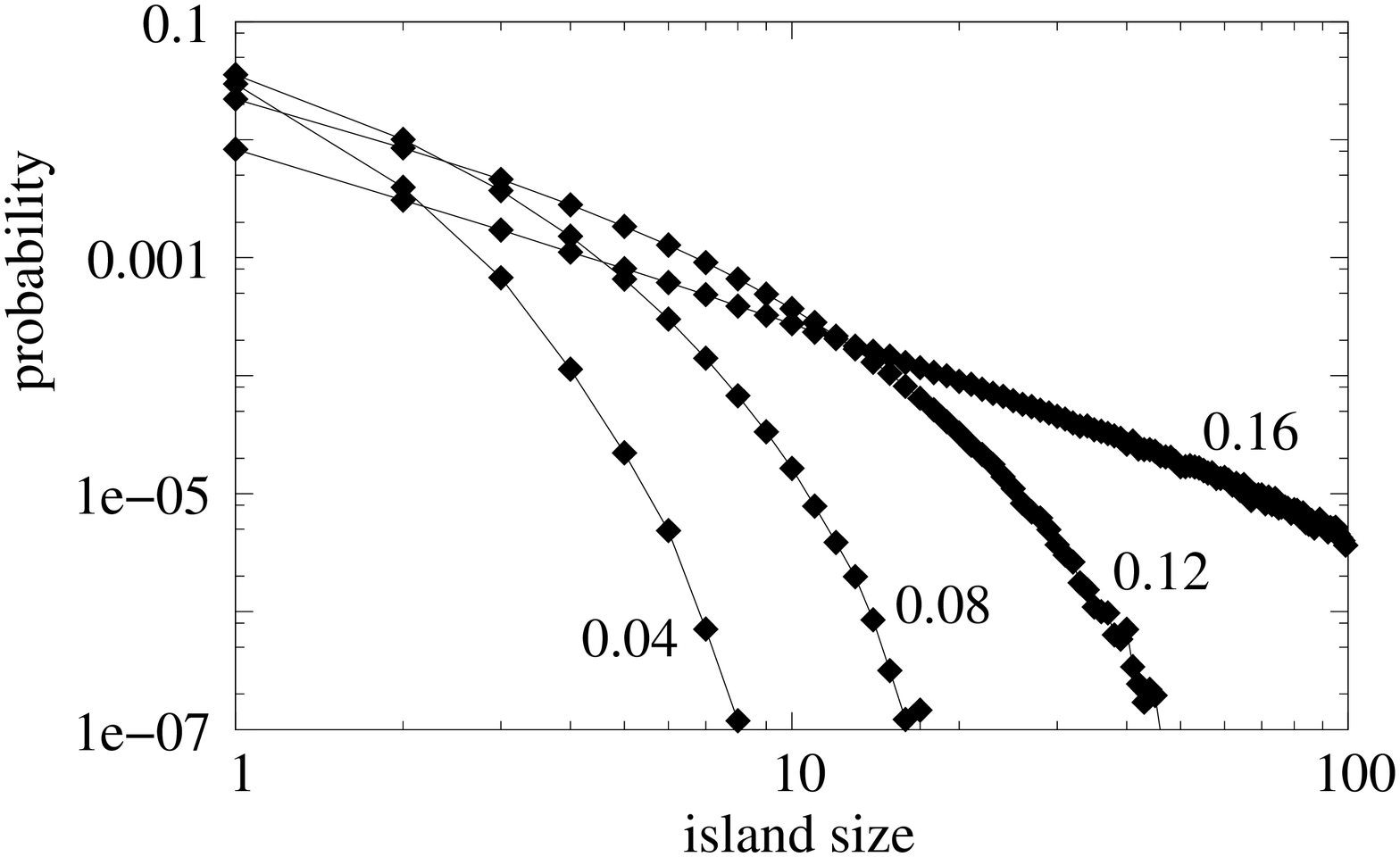,width=\hsize}
\caption{Probability per site of having an island with a certain number
  of adsorbates as a function of the number of these adsorbates. Small
  island are found with higher probability at lower coverages, whereas
  larger islands have a higher probability at higher coverages. The
  numbers next to the results stand for the coverage.
  \label{fig:clussize}}
\end{center}
\end{figure}

The islands that are formed are not static features of the adlayer. They
are also quite small at low coverages, as can be seen from
Fig.~\ref{fig:clussize}. Snapshots of the Monte Carlo simulations with
a $64\times 64$ grid show islands of about 10 to 15 oxygen atoms at a
coverage of 0.12, and 5 to 8 atoms at 0.08. At a coverage of 0.04 only
rarely more than two atoms are found together. Nevertheless, this
suffices for the structure in the LEED as shown in
Fig.~\ref{fig:hsleed}.

A remarkable aspect of the mechanism of the island formation here is the
fact that there must be repulsive interactions. If we allow two
next-nearest neighbor sites to be occupied simultaneously, we still get
island formation, but the structure observed in the LEED is then
$p(\sqrt{3}\times\sqrt{3})$. If we also allow nearest-neighbor sites to
be occupied, then no island formation takes place anymore. The reason is
that there is no entropy gain anymore by moving two adsorbates closer
together to increase the number of configurations for the other
adsorbates. Without the repulsion the number of available sites for
those other adsorbates is always equal to the number of vacant sites,
and independent of the way the two adsorbates are positioned.

A better model for O/Pt(111) than the hard-sphere model is one with
realistic values for the lateral interactions. Although DFT calculations
with cross validation indicate that the next-next-nearest neighbor
interaction can not be determined, there are other interactions that can
be obtained. We have shown that DFT results can be reproduced with an
error of only $2.6\,$kJ/mol with an adsorption energy of an isolated
oxygen atom of $-396.3\,$kJ/mol (with respect to a bare substrate and an
oxygen atom in the gas phase), a nearest-neighbor interaction of
$19.9\,$kJ/mol (positive values indicate repulsion), an next-nearest
neighbor interaction of $5.5\,$kJ/mol, and a three-particle interaction
of $6.1\,$kJ/mol that occurs if three atoms are in a row at
nearest-neighbor distances.\cite{jan02b,her06b} The last interaction can
be ignored for the coverages of interest here, because the strong
repulsion between oxygen atoms at nearest-neighbor positions prevents
even two atoms getting at these positions, let alone three. The
next-nearest neighbor interaction corresponds to a thermal energy of
about $660\,$K, so there is an appreciable probability to find two
oxygen atoms at next-nearest neighbor positions. Still, Monte Carlo
simulations with these more realistic interactions yield simulated LEED
spectra with negligible difference from those in Fig.~\ref{fig:hsleed}.

To summarize, island formation in adlayer as observed in LEED does not
need to be caused by attractive interactions between adsorbates. Island
formation can also be favored for entropic reasons, because when some
adsorbates get close together then there is more space for other
adsorbates. This extra space means that these other adsorbates can form
more different configurations and hence have a higher entropy. A
remarkable requisite for this mechanism to work is that there must be a
strong repulsion between the adsorbates at short distances. The distance
between the adsorbates in the islands is then larger than this distance
at which there is repulsion. We have shown results of this mechanism for
a hard-sphere interaction model and a model with realistic lateral
interactions for O/Pt(111), but there seems to be no reason why the
mechanism should not work in other adlayers too. Calculations of lateral
interactions seem to suggest that a strong repulsion between adsorbates
in nearest-neighbor positions is common.\cite{her06b} This means that
island formation at low coverages should be common as well. Because the
mechanism here is purely entropic and there is no energy, the island
formation is temperature independent. If there are also energetic
contributions, then the mechanism should work especially at higher
temperatures, where it may dominate interactions that favor other
adlayer structures. (Strictly speaking we have not proven that there is
no attractive interaction in O/Pt(111), but such an interaction is, as
shown by Zhdanov and Kasemo,\cite{zhd98c} too weak to be relevant.)

Ordering effects due to entropy date back at least to Onsager's hard-rod
model for liquid crystals.\cite{ons47} The depleted volume effect in
that model and the spatial effects due to the repulsive interactions
here are similar. There is an important difference however. There is a
clear distinction between degrees of freedom in the hard-rod model. In
that model the orientational entropy decreases when a nematic phase is
formed but the positional entropy increases. A similar partitioning of
degrees of freedom is found in more recent models.\cite{fre92} Here this
is not the case, and the model is simpler. Oscillations found in the
density of a gas near the wall of a microchannel could be explained with
a model with similarities to the hard-sphere model here,\cite{ned06} as
could the variation in the distribution of molecules in the channels of
a one-dimensional zeolites with variable pore diameter.\cite{sch00d}
Once again however, the model here is simpler. Moreover it is the first
model on entropic ordering in adlayers.
%
%
%

%
%
\end{document}